\documentclass[useAMS]{mn2e}

\usepackage{amssymb,amsmath,psfig,times}
\usepackage{graphicx}
\def\gsim{ \lower .75ex \hbox{$\sim$} \llap{\raise .27ex \hbox{$>$}} }
\def\lsim{ \lower .75ex\hbox{$\sim$} \llap{\raise .27ex \hbox{$<$}} }

\def\sc{Schwarzschild}
\def\beq{\begin{equation}}
\def\eeq{\end{equation}}

\def\sc{Schwarzschild}

\def\sw{{\it Swift}}
\def\fe{{\it Fermi}}
\def\ba{BATSE}

\def\ep{$E_{\rm peak}$}
\def\epop{$E^{\rm obs}_{\rm peak}-P$}
\def\epof{$E^{\rm obs}_{\rm peak}-F$}
\def\epopt{$E^{\rm obs}_{\rm peak}(t)-P(t)$}
\def\epo{$E^{\rm obs}_{\rm peak}$}

\def\pf{$P$}
\def\liso{$L_{\rm iso}$}

\def\yone{$E_{\rm peak}-L_{\rm iso}$}
\def\yonet{$E_{\rm peak}(t)-L_{\rm iso}(t)$}

\title[Short and Long GRBs]{Short and Long GRBs: same emission mechanism?}

\author[G. Ghirlanda et al.]{G. Ghirlanda
$^{1}$\thanks{E-mail:giancarlo.ghirlanda@brera.inaf.it}, G. Ghisellini$^{1}$,
L. Nava$^{2}$\\
$^{1}$INAF -- Osservatorio Astronomico di Brera, Via E. Bianchi 46, I-23807 Merate, Italy\\
$^{2}$SISSA -- via Bonomea, 265, I-34136 Trieste, Italy\\
}
\begin{document}

\date{}

%\pagerange{\pageref{firstpage}--\pageref{lastpage}} \pubyear{2002}

\maketitle

\label{firstpage}

\begin{abstract}
We study the spectral evolution on second and sub--second timescales in 11 long and 12 short Gamma Ray Bursts (GRBs) 
with peak flux $>$8.5$\times 10^{-6}$ erg/cm$^2$ s (8 keV--35 MeV) detected by the \fe\ satellite. 
The peak flux correlates with  the time--averaged  peak energy in both classes of bursts. 
The peak energy evolution, as a function of time, tracks the evolution of the flux on short timescales 
in both short and long GRBs. 
We do not find evidence of an hard--to--soft spectral evolution. 
While short GRBs have observed peak energies larger than few MeV during most of their evolution, 
long GRBs can start with a softer peak energy (of few hundreds keV) and become as hard as short ones 
(i.e. with \epo\ larger than few MeV) at the peak of their light curve.  
Six GRBs in our sample have a measured redshift. In these few cases we
 find that their correlations between the rest frame 
\ep\ and the luminosity $L_{\rm iso}$ are less scattered than their
correlations in the observer 
frame between the peak energy \epo\ and the flux \pf. 
We find that 
the rest frame \ep\ of long bursts can be as high or even larger than that of short GRBs
and that short and long GRBs follow the same \yonet\ 
correlation, despite the fact that they likely have different progenitors. 
\end{abstract}


\section{Introduction}

Since the discovery of the existence of the two classes of short and long GRBs, 
extensive studies of their prompt and afterglow properties have been finalized to outline similarities and differences. 
Short GRBs have optical and X--ray afterglows similar to those of long 
GRBs (e.g. Gehrels et al. 2008, Nysewander, Fruchter \& Pe'er 2009) but, although still in a handful 
of cases, they are typically associated with galaxies of almost all types (e.g. Berger et al. 2009) 
at preferentially lower redshifts with respect to those of long events. The leading idea is that short 
and long GRBs have different progenitors although the mechanism producing their prompt and afterglow 
emission should be similar (e.g. see Lee \& Ramirez Ruiz 2007 for a recent review). 

The comparison of the prompt $\gamma$--ray emission showed that short GRBs are harder than 
long ones, if the hardness ratio is used as a proxy of the spectral hardness (Kouveliotou et al. 1993). 
Considering the {\it time integrated} spectra of samples of short and long GRBs in the BATSE population, 
Ghirlanda et al. (2009) found that  short GRBs have a harder low energy spectral index while 
they have similar peak energy \epo\ compared to long bursts. 
Both in short and long GRBs, correlations are found between \epo\ and the fluence ($F$) and between \epo\ and the peak flux ($P$) (Lloyd \& Petrosian 1999, Ghirlanda et al. 2009, Nava et al. 2008, Nava et al. 2011b). In particular, while the \epo-$F$ correlation is different for short and long GRBs (with short events having lower fluences for the same \epo\ of long bursts), they follow the very same \epop\ correlation. Fluence and peak flux are the properties commonly used to select samples of long and short bursts in comparative studies. Different selections, based on $P$ or $F$, influence the resulting distributions of \epo\ of the selected samples and lead to different conclusions on the short-long \epo\ comparison. Selection based on the peak flux $P$ (Ghirlanda et al. 2009) results in similar \epo\ distributions for long and short GRBs, while a selection based on the fluence  $F$ leads to the conclusion that short bursts have on average larger peak energies (see Nava et al. 2011b for a detailed discussion of this effect).

Recently, Nava et al. (2011a) analysed the {\it time integrated} spectral properties of 
438 GRBs detected by the \fe\ satellite. Through this catalog, Nava et al. (2011b - N11b hereafter)
confirmed previous results, based on the \ba\ GRB population, of the existence of an \epop\ and an \epof\ correlation also in the \fe\ sample and of the different spectral properties of short and long GRBs. Thanks to the wider energy range explored by the Gamma Burst Monitor (GBM, 8keV--35MeV) on board \fe,  N11b also found that the peak energy \epo\ of \fe\ short bursts can reach very high values (confirming, with a larger population of short GRBs, what found for the three brightest short events detected by \fe\ -- Guiriec et al. 2010). Such high values of \epo\ are not so common in long bursts. However, these comparisons are typically done in  terms of observed peak energies. It is still possible that the different redshift distribution of the two classes of GRBs (e.g. Guetta \& Piran 2005) is at the origin of the difference in their \epo\ distributions.

It was known that long BATSE GRBs show a strong spectral evolution (Ford et al. 1995; Ghirlanda, Celotti \& Ghisellini 2002; Kaneko et al. 2006). This was confirmed also through time resolved spectral analysis of \sw\ bursts (Firmani et al. 2009).
The study of the spectral evolution of \fe\ long GRBs with measured redshifts 
(Ghirlanda et al. 2010, G10 hereafter) showed that a similar time resolved \epopt\ 
correlation exists within these sources. Such a positive correlation means that the observed peak energy \epo\ tracks the flux $P$ during the lifetime of the GRB.  The knowledge of $z$ for these events allowed to verify that a correlation in the rest frame 
similar to the \yone\ (so called ''Yonetoku" correlation - originally discovered considering the GRB time--integrated spectra - Yonetoku et al. 2004) exists within individual bursts.  
A similar study on a sample of \fe\ short GRBs (Ghirlanda et al. 2011, G11 hereafter) 
confirmed that also in this class there exists a time resolved \epopt\ correlation. 
These results  suggest that a mechanism could be responsible for this internal 
(i.e. time dependent) correlation between the peak energy and the flux and that this mechanism 
is similar in long and short GRBs. 

However, the sample of G10 considered 12 long \fe\ bursts with measured redshifts 
(until the end of July 2009) while the sample of G11 considered short \fe\ bursts selected on the basis of their 
peak flux and fluence (collected from the GCN circulars).
Moreover, the analysis of G10 considered only the low energy NaI detectors (8 keV--1 MeV) of the 
GBM instrument, missing the possibility to measure high \epo, while the spectral analysis of short GRBs of G11 included also the BGO detectors that extend the spectral coverage to $\sim$35 MeV. 

A direct comparison of the spectral evolution of short and long GRBs detected by \fe\ requires a 
similar selection of the two samples (see e.g. Nava et al. 2011b for a discussion of the different 
selection criteria adopted in the literature to construct short and long GRB samples) and an homogeneous 
spectral analysis over the wider energy spectral range accessible by the GBM instrument, i.e. 8 keV--35 MeV. 

%In this work we perform a  {\it time resolved} spectral analysis of two similarly selected samples 
%of short and long \fe\ GRBs (\S 2). We adopt a uniform spectral analysis method, 
%the most recent release of the \fe\ software, and we perform time resolved spectroscopy by combining 
%the NaI and BGO detectors (\S 3), i.e. in the 8 keV--35 MeV energy range. 
%Our aims are to verify the existence of the \epopt\ correlation within short and long GRBs 
%and compare them (\S 4). We also want to investigate if the apparent harder \epo\ of short \fe\ 
%GRBs with respect to long events (so far found by considering their time integrated spectra) 
%still holds when considering the time resolved spectra. We verify also (\S 5), through the few GRBs in our 
%samples with measured redshifts, if the residual difference in \epo\ in short and long GRBs could be affected by the 
%different redshift of the two populations. 

\section{The sample}
% -----------------------------------------------------
\begin{figure}
\vskip -0.4 cm
\hskip -0.5cm
\psfig{file=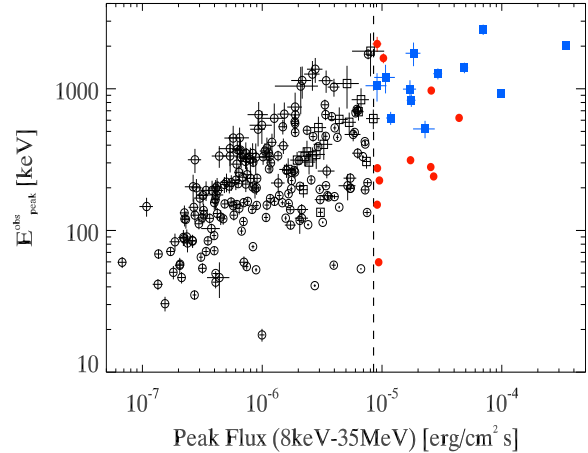,width=9cm,height=7cm}
\vskip -0.3 cm
\caption{
Observer frame peak energy versus peak flux for 235 \fe\ GRBs analyzed by N10. 
Short GRBs (31 events) are shown with squares and long GRBs (204 events) with circles. 
The sample selected for this work comprises 11 long GRBs (filled red circles) and 
12 short GRBs (filled blue square) with peak flux (8 keV--35 MeV) larger than 
8.5$\times 10^{-6}$ erg/cm$^2$ s (shown by the vertical dashed line). 
}
\label{fg1}
\end{figure}
% ------------------------------------------------------

The first compilation of the spectral properties of GRBs detected by \fe\ was recently published in 
Nava et al. (2011a, N11a hereafter). They analyzed the public data of 438 GRBs detected by the \fe\ 
GBM up to March 2010. For 318 cases they could reliably constrain the peak energy \epo\ of the 
$\nu F_{\nu}$ spectrum by analyzing the time integrated spectrum between 8 keV and 35 MeV.  
Among these 318 GRBs,  274 are long and 44 are short. Long GRBs have a typical peak energy 
\epo$\sim$160 keV and low energy photon spectral index $\alpha\sim -0.92$. Short GRBs have 
harder peak energy (\epo$\sim$490 keV) and harder low energy spectral index ($\alpha\sim -0.50$). 

N11a also analyzed the spectrum corresponding to the peak  flux of each GRB and in 235 (out of 438) 
cases they could constrain the peak energy \epo. Among these (shown in Fig. \ref{fg1})  there are 31 short 
(squares in Fig. \ref{fg1}) and 204 long (circles in Fig. \ref{fg1}) events. The peak spectrum of a burst typically 
corresponds to the time interval of the maximum of the flux in the light curve. If the spectral evolution of the 
GRB peak energy is correlated with the flux (as found independently for short and long \fe\ bursts in G10 
and G11) then the selection on the peak flux ensures that we are considering the hardest short and long 
GRBs in the analyzed sample. Moreover, the typical selection of samples of long and short GRBs present 
in the literature adopts the peak count rate which is a rather poor proxy for the actual energy flux.
The knowledge of the spectral shape for all bursts in the sample allows us to directly use the peak flux in 
physical units (8 keV--35 MeV energy flux in erg/cm$^2$/s) which accounts for the shape of the spectrum 
and for the peak energy and ensures to select at the same time the brightest and the hardest bursts both 
in the short and long population.  

We show in Fig. \ref{fg1} the two selected samples of short (filled blue squares) and long (filled red circles) 
GRBs with \pf(8 keV--35 MeV)$\ge$8.5$\times10^{-6}$ erg/cm$^2$ s. The names of the selected bursts are 
reported in Tab. \ref{tab1} together with their duration. The latter is taken from the table of N11a. 
We note that this is not the $T_{90}$ parameter typically used to characterize the GRB duration, 
but the time interval over which the spectrum was accumulated and analyzed in N11a. 
This time was selected by--eye from the light curve of each GRB as the time interval in which the signal 
was above the background. We caution that this criterion may result in a duration which is smaller 
than the $T_{90}$. However, all the short GRBs selected (except the short/hard burst 090510 
detected by the Large Area Telescope on board \fe\ - Abdo et al. 2009) have a duration much smaller than 
2 seconds which is the empirical dividing line between short and long GRBs.

%------------------------------------------------------
\begin{table}
\begin{center}
\begin{tabular}{llll}
\hline\hline
GRB   &    Duration (s) & \#  Detectors   & Max. res (ms)       \\
\hline
080802(386)   & 0.384		&	N4,N5,B0				&	32	\\
080916(009)   & 88.58		&	N4,N3,B0				&   	640	\\
081209(981)   & 0.320		&	N8,N10,B1			&	16	\\
081215(784)   & 9.468		&	N9,N10,B1			&   	128	\\
081216(531)   & 0.768		&	N8,N11,B1			&	8	\\
090227(772)   & 0.704		&	N0,N1,N2,N5,B0		&	4	\\
090228(204)   & 0.512		&	N0,N1,N2,N3,N5,B0	&	4	\\
090328(713)   & 0.192		&	N0,N1,N2,N3,N5,B0	&	16	\\
090424(592)   & 34.046	&	N8,N11,B1			&	64	\\
090510(016)   & 1.536		&	N3,N6,N7,N8,N9,B0,B1	&	8	\\
090617(208)   & 0.256		&	N3,N5,B0				&	16	\\
090618(353)   & 182.275	&	N4,N7,B0,B1			&   2048		\\
090719(063)   & 19.709	&	N7,N8,B1				&	1024		\\
090820(027)   & 32.769	&	N2,N5,B0				&	64		\\
090829(672)   & 93.186	&	N6,N10,B1			&	1024		\\
090902(462)   & 30.207	&	N0,N1,N2,B0			&	192		\\
090926(181)   & 19.709	&	N3,N7,B0,B1			&	192		\\
091012(783)   & 0.832		&	N10,N11,B1			&	32		\\
091127(976)   & 13.054	&	N7,N9,B1				&	64		\\
100116(897)   & 47.104	&	N0,N3,B0				&	1024		\\
100206(563)   & 0.256		&	N3,N5,B0				&	8		\\
100223(110)   & 0.384		&	N7,N8,B1				&	16		\\
100328(141)   & 0.768		&	N6,N11,B1			&	32		\\
\hline	
\hline		
\end{tabular}
\vskip -0.2 cm
\caption{
\fe\ GRBs selected for the time resolved spectral analysis (see Fig.\ref{fg1}). 
Detectors analyzed: N=NaI and B=BGO. The last column shows the minimum 
integration time (i.e. maximum time resolution) adopted for the time 
resolved spectral analysis. 
}
\label{tab1}
\end{center}
\end{table}

%------------------------------------------------------

% ----------------------------------------
\begin{figure}
\vskip -0.5 cm
\hskip -0.5cm
\psfig{file=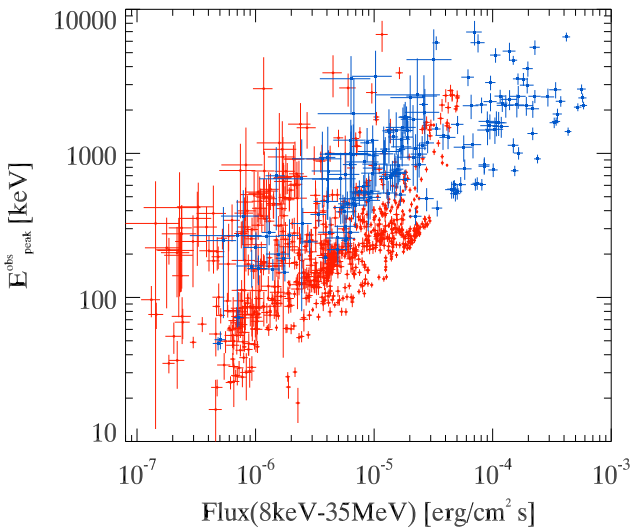,width=9.5cm,height=8cm}
\vskip -0.3 cm
\caption{
Peak energy versus flux for 749 time resolved spectra of 11 long GRBs (filled red dots) 
and 175 time resolved spectra of 12 short GRBs (filled blue squares).
}
\label{fg2}
\end{figure}
% ----------------------------------------

\section{Spectral analysis}

The GBM (Meegan et al. 2009) comprises 12 thallium sodium iodide [NaI(Tl)] and two bismuth 
germanate (BGO) scintillation detectors which cover the energy ranges  $\sim$8 keV--1 MeV 
and $\sim$300 keV--40 MeV, respectively. The GBM acquires different data types for the spectral analysis 
(Meegan et al. 2009) among which the ``TTE" event data files containing individual photons with 
time and energy tags. Here we analyze TTE data for both short and long GRBs\footnote{For the very long GRB 090618 we use the CSPEC data.} because they allow to perform 
a time resolved spectral analysis on sub--second timescales for short GRBs. CSPEC data, with 
 a maximum time resolution of 1.024s, are not suitable for time resolved 
spectral analysis of short GRBs with typical duration $<$2 s.

% ----------------------------------------
\begin{figure}
\vskip -0.4 cm
\hskip -0.5cm
\psfig{file=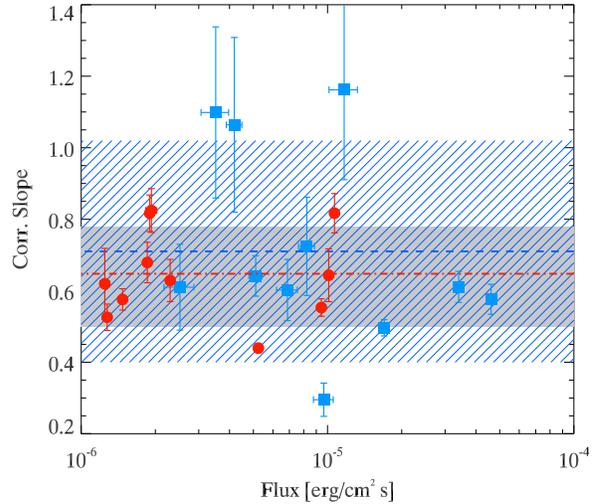,width=9cm,height=7.5cm}
\vskip -0.3 cm
\caption{
Slope $a$ of the correlation $E^{\rm obs}_{\rm peak}(t) \propto P(t)^a$ for individual bursts.
Filled red circles correspond to long GRBs, filled blue squares to short. 
The dot--dashed and dashed lines represent the average slopes for long and short GRBs, respectively. 
The solid filled and shaded regions are the 1$\sigma$ uncertainty on the average slopes. 
}
\label{fg3}
\end{figure}
% ----------------------------------------

% ----------------------------------------
\begin{figure*}
\vskip -0.4 cm
\hskip -0.5cm
\psfig{file=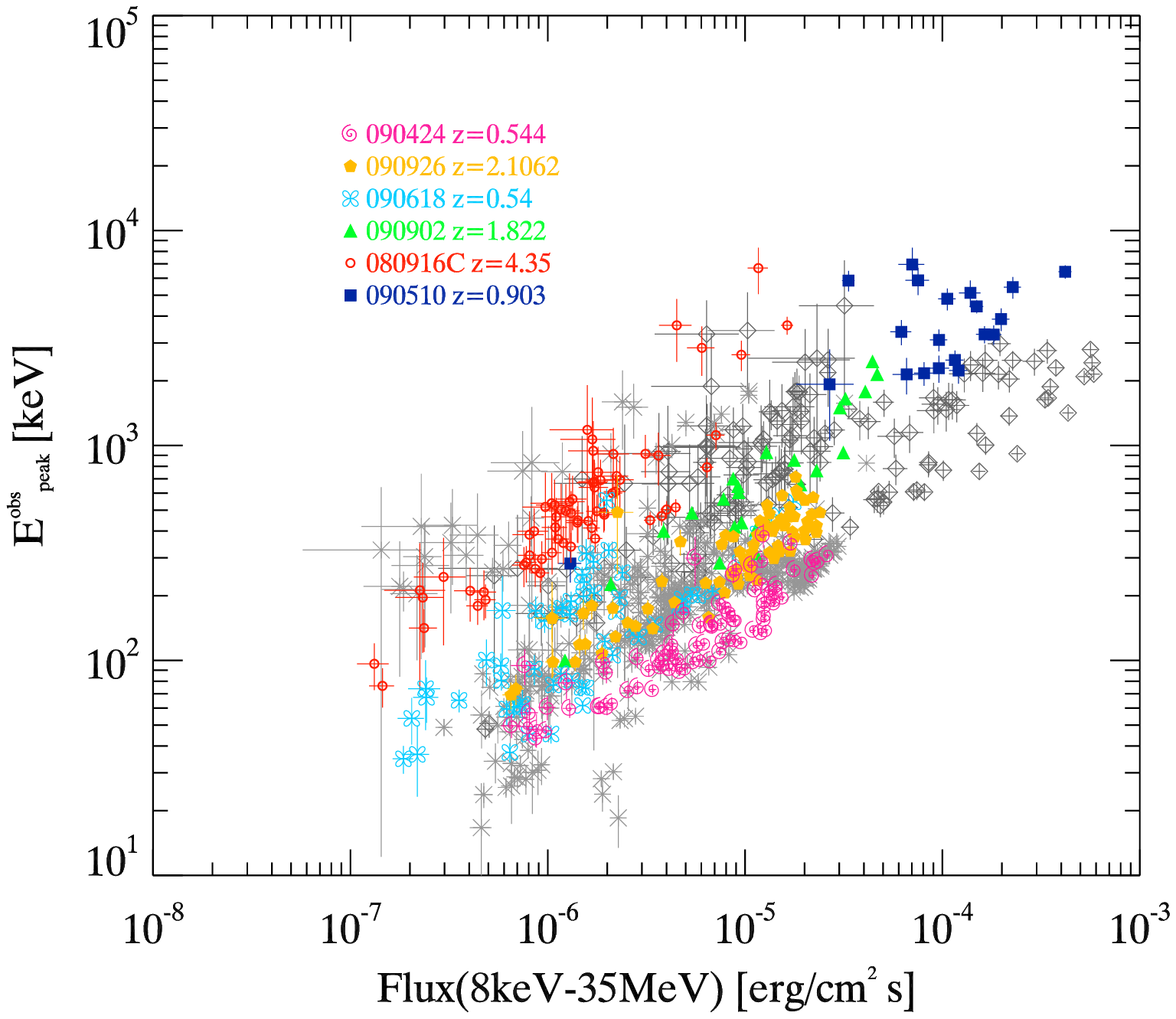,width=8cm,height=7.cm}
\psfig{file=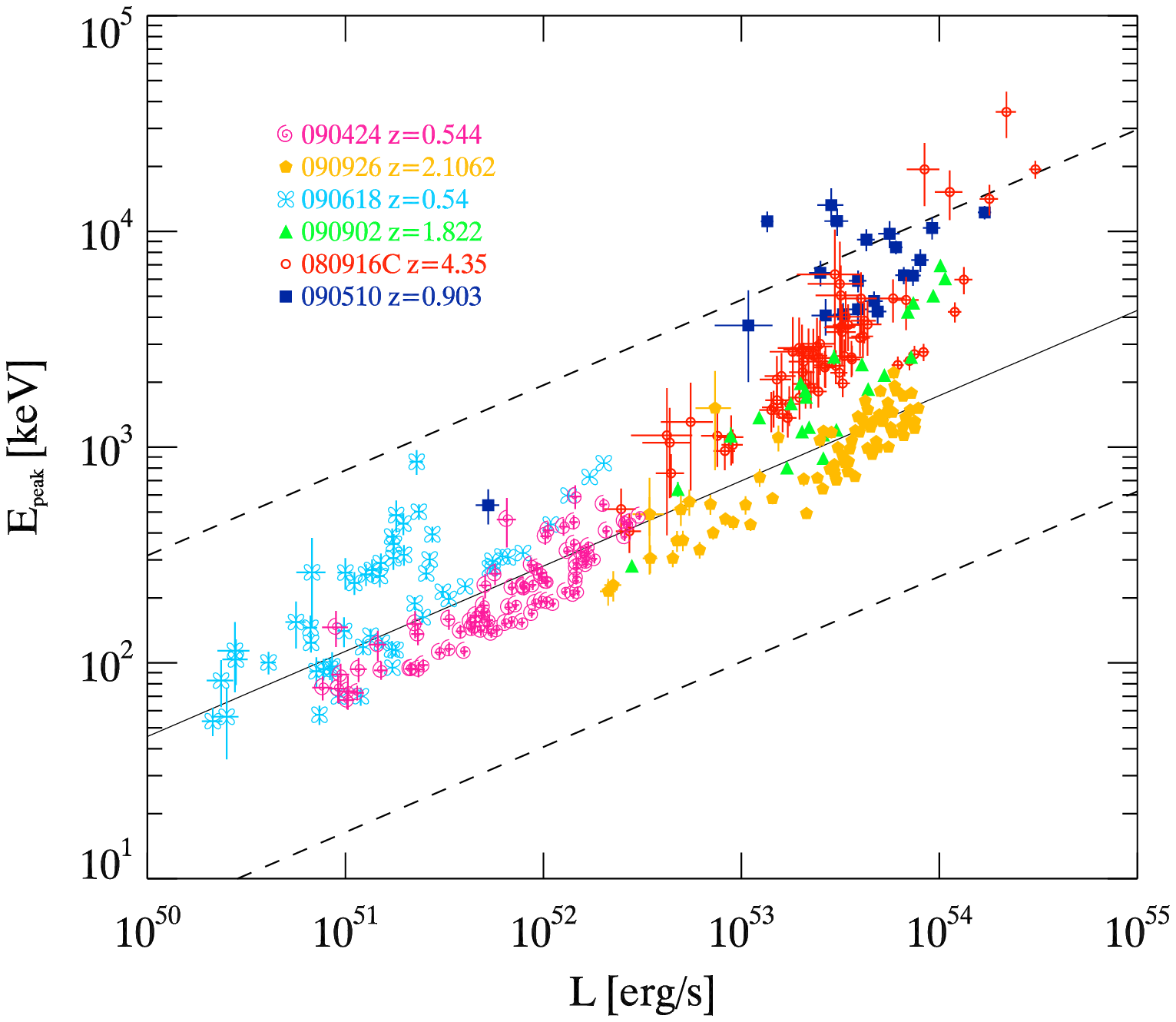,width=8cm,height=7.cm}
\vskip -0.3 cm
\caption{
Left panel: observer frame peak energy versus peak flux for the analyzed GRBs. 
The 5 long and 1 short bursts with known redshifts are marked with different 
colors. Right panel: rest frame peak energy versus luminosity for bursts with measured redshifts. 
Solid and dashed lines are the \yone\ correlation and its 3$\sigma$ scatter respectively 
(from Ghirlanda et al. 2010).
}
\label{fg4}
\end{figure*}
% ----------------------------------------

For the time resolved spectral analysis we used the recently released software 
RMFIT\footnote{http://fermi.gsfc.nasa.gov/ssc/data/analysis/user/} (\texttt{v33pr7}). 
In order to model the background spectrum for the time resolved spectral analysis, we 
selected two time intervals before and after the burst. The sequence of background spectra in the 
two selected intervals were fitted with a first order polynomial to account for 
the possible time variation of the background spectrum. Then the background spectrum was 
extrapolated to the time intervals selected for the time resolved spectroscopy of individual GRBs. 

We jointly fitted the spectra from at least 2 NaI and 1 BGO detector which 
had the largest illumination by the GRB. The inclusion of the BGO data extends the spectral coverage 
of the NaI detectors from 1 MeV to $\sim$35 MeV. Particularly bright short bursts (090227, 090228, 090328) 
were analyzed including more than 2 NaI (and for 090510 both BGO detectors) as done by Guirec et al. (2010). 
We performed a time resolved spectral analysis by progressively reducing the time 
bin duration. For the brightest GRBs of our sample (i.e. GRB 090510 and GRB 090227 both of the short class) 
we performed a time resolved spectral analysis down to 4 ms. In general the time resolution within a GRB is higher near the 
peak and lower along the tails of the burst. This was determined by the requirement that all the parameters of the 
fitted model to each time resolved spectrum could be constrained (i.e. their associated statistical errors determined 
with an accuracy $<$ 100 \%).  In order  to have a reasonable number of time resolved spectra and to make the 
plot of Fig. \ref{fg2} readable, we limited the time resolution to 1024 or 2048 ms for particularly long GRBs. 
In Tab. \ref{tab1} (last column) we report for each GRB analyzed the shortest time bin on which their 
spectrum was analyzed.

The model adopted for the time resolved spectral analysis is a power law with an exponential 
cutoff whose free parameters are the low energy spectral index $\alpha$, the peak energy 
\epo\ (i.e the peak of the $\nu F_{\nu}$ spectrum), and the normalization. 
The choice of this model is motivated by the fact that at high energies the 
response of the BGO rapidly decreases for increasing energy so that it is hard, 
in single time resolved spectra,  to constrain the possible presence of a power law component of e.g. 
the Band function (Band et al. 1993) which is instead typically fitted to the time integrated 
spectra of long GRBs (see also N11b for a discussion on the model choice). 
The choice of the same model for the time resolved analysis of short and long GRB spectra 
ensures that if the model is introducing any systematic effect in the estimate of some of its 
spectral parameters, this is common to all the analyzed spectra.  
Note that if short and long GRBs spectra have intrinsically different shapes (e.g. 
a Band model for long and a cutoff--powerlaw for short) our choice would introduce a systematic 
effect, i.e. an overestimate of the peak energy of long GRBs (e.g. Kaneko et al. 2006). Still it is hard, in time resolved 
spectra, to distinguish between a cutoff--powerlaw model and a Band function given the lower response 
of the instrument at high energies and the low fluxes of single time resolved spectra.

\section{Results}

By considering the evolution of the peak energy \epo\ of the $\nu F_{\nu}$ spectrum and the energy 
flux \pf\ (integrated over the 8 keV--35 MeV energy range), we find that in most bursts 
there is a {\it tracking} pattern. This is shown by the strong correlation between \epo\ and \pf\ 
shown in Fig. \ref{fg2}. This correlation is shown for both short (filled blue squares) and long 
(filled red circles) GRBs and extends over four and three decades in \pf\ and \epo\ respectively. 
By considering long and short GRBs separately, the Spearman correlation coefficient is $r$=0.54   
(with a chance probability of $10^{-58}$) and $r$=0.74 (with a chance probability of $10^{-31}$) 
for long and short GRBs, respectively. 

We also fitted the \epopt\ correlation within the 12 short and 11 long GRBs considered individually. 
The slopes $a$ of the correlation $E^{\rm obs}_{\rm peak}(t) \propto P(t)^a$ for individual bursts are shown
in Fig. \ref{fg3} with different symbols for long (filled red circles) and short (filled blue squares) GRBs. 
Fig. \ref{fg3} also shows the slope (dashed and dot--dashed lines) obtained by averaging the slopes 
of short and long GRBs. The slopes of the \epopt\ correlation in the sample of short GRBs are more scattered, 
with also one case (i.e. GRB 081209) having a very flat correlation. In general the correlation slopes of short 
and long GRBs are similar (as shown by their average values) and consistent with a typical slope of 0.5 also 
found within the sample of \fe\ GRBs by considering the time integrated spectra (N11b). Moreover, this is also 
consistent with  the slope of  the ``classical" Yonetoku correlation, \yone, (which is computed in the rest frame 
for bursts with measured redshifts) obtained considering the time integrated spectra of long and short 
GRBs (see e.g. G10 for a recent compilation of this correlation).

Short and long GRBs follow the same correlation but they are slightly 
displaced in Fig. \ref{fg2}, with short GRBs having larger fluxes and 
peak energies with respect to long GRBs, similarly to what found with the time integrated 
spectral properties of \fe\ and \ba\ long and short GRBs (N11b). However, among the sample of 
long GRBs there is GRB 080916C whose peak energy \epo\ reaches values of few tens of 
MeV at the peak and becomes as hard as short GRBs (better shown in Fig. \ref{fg4} - left panel). 

The quantities plotted in Fig. \ref{fg2} are in the observer frame.
Since the average redshift of long GRBs is larger than that 
of short ones, the displacement shown by Fig. \ref{fg2} between short and long
is probably reduced in the rest frame. In our samples of bursts there are  6 bursts 
with known $z$. Among these one is the short GRB 090510 at z=0.903. 
In Fig. \ref{fg4} we highlight with colors and different symbols these bursts in the observer 
frame (left panel) and in the rest frame (right panel). 
We notice that when transforming from the observer frame \epopt\ to the rest frame \yonet\ 
the scatter of the correlations of these 6 bursts is evidently reduced. 
Moreover, in the rest frame (right panel in Fig.\ref{fg4}) the long 
GRB 080916C (at $z=4.35$) has the largest peak energies (i.e. larger than the short GRB 090510). 

\section{Discussion and Conclusions}

%In this work we selected with the same criterion two representative samples 
%of bright long and short \fe\ bursts (Fig. \ref{fg1}). We considered the 23 GRBs with peak flux 
%larger than $8.5\times10^{-6}$ erg/cm$^2$ s (integrated in the 8 keV--35 MeV energy range) 
%in the \fe\ GRB spectral catalog of N11a. Of these 23 GRBs, 11 are long and 12 short.

Compared to previous analysis (G10, G11 and Guiriec et al. 2010) that considered separately 
short and long GRBs selected for their peak flux or fluence, here we ensure to have 
a complete sample down to our selection threshold. Moreover, we performed a homogeneous 
time resolved analysis of these bursts over the 8 keV -- 35 MeV energy range by combining 
the NaI and BGO detectors of the GBM instrument on board \fe. 

We find that both short and long GRBs follow a strong \epopt\ correlation in the observer  
frame (Fig. \ref{fg2}).  Considering single bursts, the slopes $a$ of the 
$E^{\rm obs}_{\rm peak}(t) \propto P(t)^a$ correlation are consistent with a $a\sim$0.5 
(with short GRBs having a larger scatter of the values of $a$ than long ones, see Fig. \ref{fg3}).

The existence of a \epopt\ correlation in short and long GRBs indicates that the typical spectral 
evolution in both classes of events is {\it tracking} rather than hard--to--soft. 
This result is different from the dominating hard--to--soft spectral evolution found in a 
sample of \ba\ bursts (Hakkila \& Preece 2011) and in agreement with past studies based on 
the time resolved spectral modeling of bright \ba\ GRBs (Ford et al. 1996, Ghirlanda et al. 2003). 
A possible reason could be that here we model the single time resolved spectra in each GRB 
with a spectral model (the cutoff--power law function) and use the resulting \epo\ as a proxy 
of the GRB hardness instead of the count hardness ratio as done in Hakkila \& Preece (2011).
 
Only 6 GRBs in our sample have measured reshifts and, among these, we have only one short GRB.
Fig. \ref{fg4} shows that in the \yonet\ plane these bursts have a reduced scatter with respect to the 
\epopt\ plane. Moreover, in the rest frame plane, long bursts can reach values of $E_{\rm peak}$ as large as
those of the only short one with known $z$, contrary to what occurs for the observed values (see e.g. Guirec et al. 2010).
If these results, now based on a very limited number of events, would be confirmed by the addition of short and long GRBs 
with measured redshifts with a time resolved spectral analysis, it could indicate that 
short bursts {\it appear} harder (i.e. larger $E_{\rm peak}^{\rm obs}$), than long GRBs,
only because of their different average redshift. 
Interestingly, the 6 GRBs with measured redshift have individual time resolved \yonet\ correlations which are 
all consistent with the \yone\ correlation (solid line in Fig.\ref{fg3}) found with the time integrated 
spectra of a much larger sample of GRBs.

Several interpretations of the \yone\ correlations have been proposed so far: (a) kinematic 
interpretations in which the link between \ep\ and \liso\ is established by the configuration 
of the emission region (Yamazaki et al. 2004; Nakamura 2000; Kumar \& Piran 2000; Toma et al. 2005; 
Eichler \& Levinson 2004); (b) radiative interpretations in which it is the emission mechanism of 
the prompt phase to link \ep\ and \liso, as in the case of a spectrum dominated by a thermal component 
(Ryde et al. 2006; Thompson, Meszaros \& Rees 2007) or in the case of magnetic 
reconnection (Giannios \& Spruit, 2007) or (c) both radiative and geometric effects (Lazzati et al. 2010) . All these interpretations however considered the time 
integrated \yone\ correlation found in the population of long GRBs. Here we have shown that two 
new ingredients should be considered: (1) that a time resolved \yonet\ correlation exists within 
individual GRBs and that (2) this is similar in the population of long and short GRBs. 
Among the above mentioned models, Giannios \& Spruit (2007) predict that a time resolved \yonet\ 
correlation should exists within single GRBs and that more luminous GRBs should have a larger baryon loading. 

Our results indicate that both short and long GRB have in common a similar physical mechanism 
that link the flux and the peak energy within individual bursts as a function of time. 
If short and long GRBs have a different progenitor (the merger of two compact objects and 
the death of a massive star respectively) the similar \yonet\ correlation found in both 
populations should arise from the similar emission mechanisms
(independent of the type of progenitor). 
%This result is also a strong 
%argument against claims that the \yone\ correlation is due to instrumental selection effects 
%(e.g. Butler et al. 2007, 2009).  
We have shown that within individual GRBs (when considering time resolved spectra) there is 
strong correlation between the peak energy and the peak flux (or luminosity for the few events with measured redshifts) and 
this correlation has the same slope of the correlation defined among different GRBs (by considering their time integrated spectral properties - 
either the peak energy--peak flux correlation or the peak energy--luminosity Yonetoku correlation).  The found 
\epopt\ correlation cannot be due to the detector threshold, and its coincidence with the \epop\ correlation found among different GRBs suggests that 
the latter is not due to instrumental selection effects but should have instead a physical origin.

\section*{Acknowledgments}
We acknowledge ASI (I/088/06/0) and a 2010 PRIN--INAF grant for
financial support. This research has made use of the data obtained through
the High Energy Astrophysics Science Archive Research Center Online Service,
provided by the NASA/Goddard Space Flight Center. LN thank  
the Brera Observatory for the kind hospitality during the completion of this work. 
We are grateful to E. Nakar for useful comments that improved the manuscript.


\begin{thebibliography}{}






\bibitem{} Abdo, A. A., Ackermann, M., Ajello, M., et al., 2009, Nature, 462, 331
\bibitem{} Band, D., Matteson, J., Ford, L. et al. 1993, ApJ, 413, 281
\bibitem{} Berger, E. 2006, ``Gamma-Ray Bursts in the Swift Era'' conference  proceeding, AIP, 836, 33
\bibitem{} Berger, E., 2009, ApJ, 690, 231
\bibitem{} Butler, N. R., Kocevski, D., Bloom, J. S., \& Curtis, J. L. 2007, ApJ, 671, 656 
\bibitem{} Butler, N. R., Kocevski, D., \& Bloom, J. S. 2009, ApJ, 694, 76 
\bibitem{} Eichler, D., \& Levinson, A. 2005, ApJ, 635, 1182
\bibitem{} Firmani, C.; Cabrera, J. I.; Avila-Reese, V., et al., 2009, MNRAS, 393, 1209
\bibitem{} Ford, L. A.; Band, D. L.; Matteson, J. L., 1995, ApJ, 439, 307
\bibitem{} Gehrels, N.; Barthelmy, S. D.; Burrows, D. N. et al., 2008, ApJ, 689, 1161
\bibitem{} Ghirlanda, G., Celotti, A., \& Ghisellini, G. 2002, A\&A, 393, 409
\bibitem{} Ghirlanda, G., Ghisellini, G. \& Celotti, A., 2004, A\&A, 422,L55 
%\bibitem{} Ghirlanda, G., Ghisellini, G., Firmani, C. 2005, MNRAS, 361, L10
\bibitem{} Ghirlanda, G., Nava, L., Ghisellini, G., Firmani, C., Cabrera, J.I., 2008, MNRAS, 387, 319
\bibitem{} Ghirlanda, G.; Nava, L.; Ghisellini, G.; Celotti, A.; Firmani, C.,  2009, A\&A, 496, 585
\bibitem{} Ghirlanda, G.; Nava, L.; Ghisellini G., 2010, A\&A, 511, 43
\bibitem{} Giannios, D., \& Spruit, H. C. 2007, A\&A, 469, 1
\bibitem{} Guetta, D. \& PIran, T., 2005, A\&A, 435, 421
\bibitem{} Guiriec, S.; Briggs, M. S.; Connaugthon, V., et al.,  2010, ApJ, 725, 225
\bibitem{} Hakkila, J., \& Preece, R., 2011, ApJ subm., arXiv:1103.5434 
\bibitem{} Kaneko, Y.; Preece, R. D.; Briggs, M. S.; 2006, ApJS, 166, 298 
\bibitem{} Kouveliotou, C., Meegan, C. A.. Fishman, G. J. et al. 1993, ApJ,  413, L101 
\bibitem{} Lazzati, D., Morsony, B. J., Begelman, M. C., 2011, ApJ, 732, 34
\bibitem{} Lee, W. H. and Ramirez-Ruiz, E. 2007, NJPh, 9, 17
\bibitem{} Lloyd, N. \& Petrosian, V., 1999, ApJ, 511, 550
%\bibitem{} Levinson, A., \& Eichler, D. 2005, ApJ, 629, L13 
%\bibitem{} Lu, R.--J., Hou, S.--J., Liang, En-We, 2010, ApJ, 720, 1146
%\bibitem{} McBreen, S., et al., 2008, GCN 8680 Ê Ê Ê%081216
\bibitem{} Nava, L., Ghirlanda, G., Ghisellini, G, Firmani, C. 2008, MNRAS, 391, 639
\bibitem{} Nava, L., Ghirlanda, G., Ghisellini, Celotti, A., 2011a, A\&A 530, A21 (N11a)
\bibitem{} Nava, L., Ghirlanda, G., Ghisellini, Celotti, A., 2011b, MNRAS in press., arXiv:1004.1410 (N11b)
\bibitem{} Nysewander, M.; Fruchter, A. S.; Pe'er, A., 2009, ApJ, 701, 824
%\bibitem{} Panaitescu, A. 2009, MNRAS, 393, 1010
%\bibitem{} Rau, A., et al., 2009, GCN 9353 Ê %090510 redshift
%\bibitem{} Rees, M., \& Meszaros, P. 2005, ApJ, 628, 847
\bibitem{} Ryde, F., Bjornsson, C., Kaneko, Y., et al. 2006, ApJ, 652, 1400
%\bibitem{} Shahmoradi, A., \& Nemiro?, R. J. 2009, [arXiv:0904.1464]
\bibitem{} Toma, K., Yamazaki, R., \& Nakamura, T. 2005, ApJ, 635, 481 
%\bibitem{} Thompson, C. 2006, ApJ, 651, 333
\bibitem{} Thompson, C., Meszaros, P., \& Rees M. J. 2007, ApJ, 666, 1012
\bibitem{} Yamazaki, R., Ioka, K., \& Nakamura, T. 2004, ApJ, 606, L33
\bibitem{} Yonetoku, D., Murakami, T., Nakamura, T. et al. 2004, ApJ, 609, 935

\end{thebibliography}
\end{document}